# Trends of digitalization and adoption of big data & analytics among UK SMEs: Analysis and lessons drawn from a case study of 53 SMEs

Muhidin Mohamed, Philip Weber
School of Engineering and Applied Science
Aston University, Birmingham, B4 7ET, UK
m.mohamed10@aston.ac.uk; p.weber1@aston.ac.uk

*Abstract—* **Small and Medium Enterprises (SMEs) now generate digital data at an unprecedented rate from online transactions, social media marketing and associated customer interactions, online product/service reviews and feedback, clinical diagnosis, Internet of Things (IoT) sensors, and production processes. All these forms of data can be transformed into monetary value if put into a proper data value chain. This requires both skills and IT investments for the long-term benefit of businesses. However, such spending is beyond the capacity of most SMEs due to their limited resources and restricted access to finance. This paper presents lessons learned from a case study of 53 UK SMEs, mostly from the West Midlands region of England, supported as part of a 3-year ERDF[1] project – Big Data Corridor[2] – in the areas of big data management, analytics and related IT issues. Based on our study's sample companies, several perspectives including digital technology trends, challenges facing the UK SMEs, and the state of their adoption in data analytics and big data, are presented.**

## 1. INTRODUCTION

Data is now arguably defined as the *new oil* of this digital era [1], and about 90% of the massive amount of data we have today has been generated in the last two years alone [2]. This indicates how valuable data is for businesses, and the extent to which raw data without analysis, like unrefined crude oil, will be of little use for the growth of SMEs.

The World Bank estimates that businesses formally categorised as Small or Medium Enterprises (SMEs) contribute up to 60% to the total employment and about 40% to the GDP in emerging economies[3]. The figures are even higher in the EU where 99% of businesses are classified as SMEs, employing about 66% of the total workforce [3]. An SME is generally defined as a company having less than 250 employees and a turnover not exceeding €50 million. This covers sole trader, micro, small and medium-sized businesses. Digitalization and the use of data analytics offer such businesses new opportunities, such as marketing optimization, forecasting the demand for their products and services, and staying ahead in the competition for customer acquisition and retention. Governmental institutions, including the EU, recognize the importance of empowering SMEs to join the digital revolution and monetize their data, as reflected in the proportion of the Horizon 2020 EU projects for data-related innovation, big data, and data science [4].

In the UK, SMEs account for 99.9% of all private sector businesses, creating 60% of the employment[4], and producing about 52% of all private sector turnover [5]. This shows the important role that SMEs play in the UK economy and that any means of helping their rapid growth could be a huge boost to the country's economy. This view is substantiated by a survey study of 500 UK companies which found a positive correlation between the use of data, and business performance and productivity: top data-using firms are 13% more productive than those in the bottom quartile [6]. Despite this, the adoption of big data analytics by UK SMEs is less than 1% [7-8]. Even so, the interest and understanding of data-driven decision-making is gaining ground among UK businesses. Figure 2 illustrates the Google trends for the terms: *big data*, *data analytics*, and *big data analytics* in the UK for the last ten years. It is clear that the interest in the use of data, big data and their analytics has grown over time. The popularity of *data analytics* continues to grow, unlike *big data* and its *analytics* which slowed down in 2018, although the trend of *big data* analytics picked up in 2019.

In this paper we summarize the lessons learned during a 3-year ERDF project aimed at assisting SMEs in the areas of data management, analytics, big data, and related digital issues. We also briefly analyse the challenges facing SMEs who want to make use of data-driven innovation and decision-making, highlighting areas where small businesses are most in need of digital technology support. Due to non-disclosure agreements and SMEs' business confidentiality, we do not include details of support given to individual SMEs nor any associated data. The remainder of the paper is organized as follows: Section 2 recaps concepts of *big data* and *data analytics*, then in Section 3 we briefly analyse SMEs' data and digital technology trends. Section 4 presents related discussion and the key lessons learned while supporting and collaborating with SMEs. We provide a summary and conclusion in Section 5.

---

[1] European Regional Development Fund.
[2] An ERDF funded 3-year project from 2016 to 2019: https://bigdatacorridor.com/.
[3] https://www.worldbank.org/en/topic/smefinance.
[4] This is accounting for the fact that 75% of SMEs are non-employing businesses, i.e. self-employed or sole traders.

Table 1: The five characteristic attributes (Vs) of big data

| | |
|---|---|
| **Volume** | Describes the magnitude of data pouring from a variety of sources, including CCTV, Internet of Things (IoT) transport sensors, social media, online transactions, and electronic health records, all making the handling of such data impossible using conventional technologies. |
| **Variety** | Relates to the different data types (e.g., video, images, text, etc.), structural heterogeneity (structured, semi-structured and unstructured), and the different sources that generate these kinds of data. |
| **Velocity** | This refers to the rate at which data is generated online and the processing speed required for its analysis, for example, we send 188 million emails and make about 3.8 million searches on Google alone every minute[5]. |
| **Veracity** | Relates to the accuracy and trustworthiness of the data before storing and mining it for insights and business intelligence, which is becoming increasingly important in this age of rising cybercrime and misinformation. |
| **Value** | Refers to the usefulness, monetization, and cost-benefit analysis of data, as data however enormous in size, will be of little use and benefit if not put into a proper data value chain. |

## 2. BIG DATA AND DATA ANALYTICS

The terms *big data* and *data analytics* are closely related and both underpin the broader field of data science. Although there is no commonly agreed definition for *big data*, the term is generally used to mean large and complex data, which cannot be handled with conventional data storage and processing tools. Gandom & Haider [9] provide several definitions of big data based on an online survey of 154 global executives. These are based on the source of and processes around data – explosion of new (largely internet-related) data generating sources, growth in transactions, new technologies for processing and analysing data, and regulatory requirements for storage and retention.

In addition, big data is widely described along at least 3 [10] 'V' dimensions (Volume, Variety and Velocity) to which are commonly added Variety and Veracity giving rise to five Vs (Table 1). Additional dimensions including Variability and Visualisation are sometimes added. On the other hand, *data analytics* is the technical process of transforming raw data into meaningful insights for better decision making. This includes text analytics, e.g., summarization, since unstructured text forms the highest percentage of current big data [11-14].

There are four distinct types of data analytics, namely, descriptive, diagnostic, predictive, and prescriptive, based on the level of analytic maturity in a business [15]. These four stages and associated factors (e.g., value, complexity and analytic questions) are illustrated in Figure 1.

---
[5] https://lorilewismedia.com/

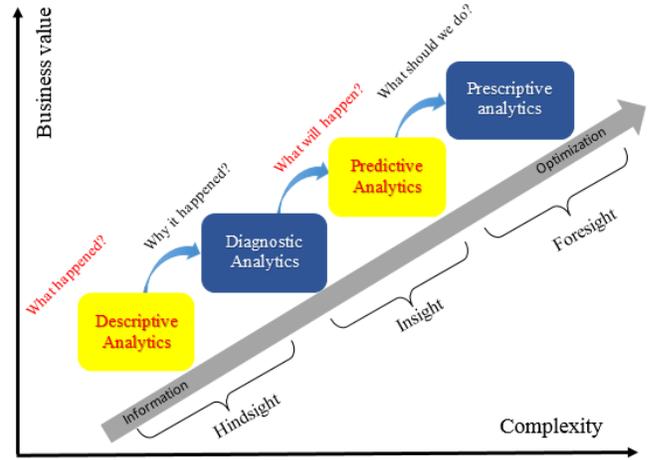

Fig. 1: The four types of data analytics, and the value, difficulty, and the questions they answer [5].

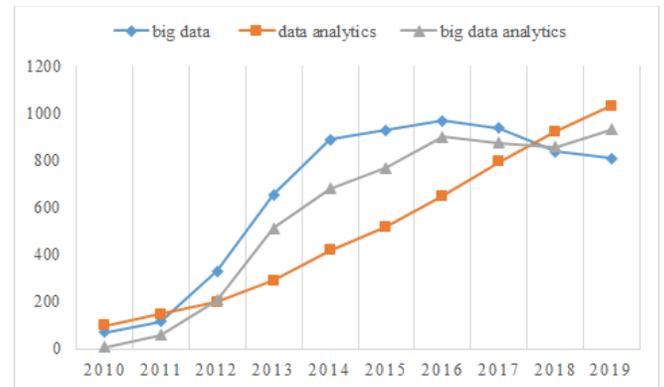

Fig. 2: Google trends for: big data, data analytics, and big data analytics in the UK for the last 10 years.

Adoption of data analytics can provide SMEs with a competitive advantage, for example, enabling them to develop a new data-based business optimization that could lead to increased revenue and profit.

## 3. A BRIEF ANALYSIS OF SMEs' DATA & DIGITAL TRENDS FROM DIFFERENT PERSPECTIVES

For this analysis and discussion, we use a case study of 53 companies, mostly from the West Midlands region of England. These SMEs were selected due to their support being delivered by the Aston University research group, one of the Big Data Corridor (BDC) project partners, from June 2017 to August 2019. Figure 3 shows the BDC project support lifecycle for SMEs. Statistically speaking, the firms employed from one person (the owner) to over 140 employees with a turnover ranging from £10,000 to over £5 million. Table 2 shows the top 10 SME business sectors, and ranges of their staff numbers, turnover figures, and the duration of support in months. The principal objective of this project was to help SMEs recognize the hidden values of data from various data sources including photonic-based light and temperature sensors, and the IoT. Our main aim was also to enable SMEs to derive meaningful insights from closed (belonging to the business) and open (publicly available) data through analytics. To this end, the project has helped SMEs in the following:

  1) providing training and skills-upgrading seminars and workshops in data collection, storage and analysis, and data-related technologies, such as photonics, IoT, and sensors;
  2) giving SMEs face-to-face technical consultancy on system design and development, data management and analytics, and offering them an opportunity of further research collaboration; and
  3) helping SMEs understand and identify data-driven services and products using both their own, and open data.

Table 2: Top ten business sectors among the 53 SMEs: supported from June 2017 to August 2019.

| No. | SME sector and proportion | No. of staff | Turnover | Support period |
|---|---|---|---|---|
| 1 | Consultancy (15.1%) | 1 – 15 | £ 50 – £ 650 k | 2 – 4 months |
| 2 | Technology (13.2%) | 1 – 75 | £ 10 k – £ 5.5 m | 2 – 4 months |
| 3 | Education and training (11.3%) | 1 – 20 | £ 10 k – £ 700 k | 2 – 8 months |
| 4 | Marketing (9.4% ) | 1 – 10 | £ 30 k – £ 300 k | 2 – 6 months |
| 5 | Manufacturing (7.5%) | 1 – 60 | £ 20 k – £ 900 k | 2 – 5 months |
| 6 | Travel and tourism (7.5%) | 1 – 10 | £ 35 k – £ 700 k | 2 – 4 months |
| 7 | Recruitment/employment (7.5%) | 1 – 60 | £ 40 k – £ 2.5 m | 1 – 4 months |
| 8 | Food (5.7%) | 1 – 45 | > £ 1.5 m | 3 – 5 months |
| 9 | Property management (3.8%) | 1 – 5 | £ 30 – £ 300 k | 2 – 4 months |
| 10 | Legal solicitors (3.8%) | 2 – 40 | £ 400 k < | 4 months |

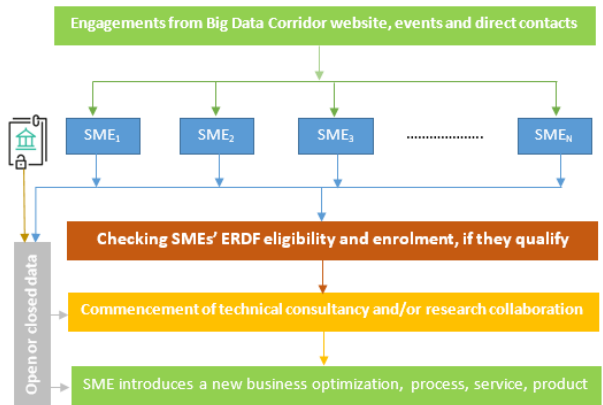

Fig. 4: BDC project support lifecycle for SMEs.

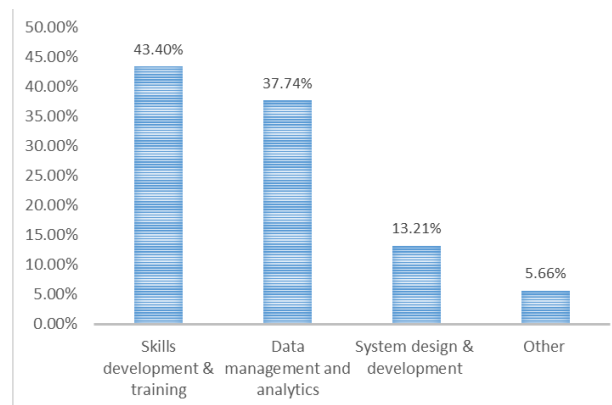

Fig. 3: % of the type of assistance given to SMEs.

The 53 beneficiary SMEs of this study were drawn from 18 different business sectors. Table 2 shows the top ten business sectors represented, together comprising about 85% of the total supported SMEs. The highest proportions of the SMEs were from the consultancy (15.1%), technology (13.2%), and education & training (11.3%) sectors. This suggests that service and technology industries significantly rely on the use of data-driven approaches for their business, as compared to other merchandising SMEs. The sectors of the remaining SMEs included recycling, fire safety, event management, and accounting.

In Figure 4, we show an analysis of the digital technology support categories within which assistance was provided and the proportions of SMEs that sought each type of assistance. Training and skills enhancement, in the form of workshops and seminars, was the most frequently accessed type of assistance, representing 43.4% of which 18% related to data use. Next, SMEs supported in data management and analytics, at both descriptive and predictive levels, constituted 37.7%, followed by photonics-related system design and development which amounted to 13.2%. From our analysis, we found four main analytic aspects in which SMEs require assistance, namely;

  1) analysing customer purchasing patterns and profiles using past business data with the aim of optimizing existing marketing strategies;
  2) summarizing data to extract useful insights in order to improve service and/or product quality;
  3) examining the impact of services such as studying the correlations between training and education programs and participants' achieved outcomes; and

Table 3: An illustration of SMEs' business sectors and the digital technology issues they have been supported with.

|  | Training & Skills development | Data management and analytics | Electronic System development | Other |
|---|---|---|---|---|
| Travel and tourism | 1 | 3 |  |  |
| Education and training | 1 | 5 |  |  |
| Recruitment | 2 | 2 |  |  |
| Property management |  | 2 |  |  |
| Manufacturing |  | 1 | 3 |  |
| Marketing | 3 | 1 |  | 1 |
| Technology | 4 | 2 | 1 |  |
| Food |  | 1 | 2 |  |
| Legal solicitors | 2 |  |  |  |
| Consultancy | 7 | 1 |  |  |

4) *identifying potential markets for SMEs' products and services based on analytics of relevant open datasets.*

The most common data formats among SMEs, seen during the analytics support, ranged from simple-format CSV files and Excel spreadsheets to complex structured data in JSON[6] format. In most descriptive and in all predictive analytics support, we used open-source tools, such as Microsoft Power BI and open source RStudio environments, with the aim of encouraging SMEs to continue to use them as they are free to use and develop.

In general, we observed that most businesses collect and store some sort of data but require additional skills to conduct advanced data analysis and produce useful statistics and correlations that can support them in making strategic decisions. Data management, including retrieval, processing, and storing data from external websites and third-party cloud-based applications, was found to be in high demand. This was also the case for essential mechanisms for securing data and compliance with GDPR. Overall we noted that small businesses lag behind their larger counterparts when it comes to using their data for deeper statistical analysis and predictive modelling – as indicated by their low adoption of predictive analytics (approx. 5% in this data).

Next we illustrate in Figure 5 a breakdown of the supported SMEs based on the level of support. As shown, 62% of the SMEs received "business assist", a time-limited short assistance including attending training workshops and/or one-to-one consultancy and technical support. The "\enterprise assistance" taken up by the other 38% has been achieved through longer research collaborations, both necessitated by and made possible by the scale and complexity of the problems and opportunities faced by these SMEs. A typical research collaboration under this project includes analysing data for the SMEs to extract meaningful insights, developing bespoke data analytic tools and data models using machine learning, and supporting the SME with the design, development and demonstration of electronic data sensors. Conducting research and transferring its findings and training in any necessary skills formed part of the collaboration. The right-hand chart shows that of the 38% "enterprise assistance" collaborations, 32% resulted in improvement of an SME's business operations and its marketing, whereas the remaining 6% benefited either the SME business process or their marketing. We can conclude from this that the scale of digital technology and data analytic issues facing UK small businesses is not trivial, requiring long-term and sustained investment and support.

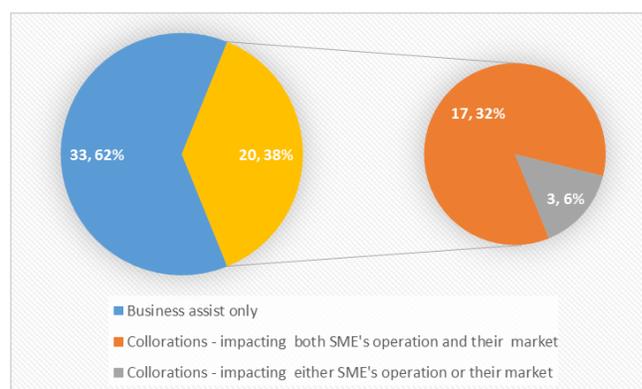

Fig. 5: Breakdown of the beneficiary SMEs by the level of support receive, showing size of the "business assist" and "enterprise assistance" cohorts (left, by number of SMEs and percentage), and expanding the "enterprise assist" cohort (right).

One final observed trend to report from this study is the relationship between business sectors and digital technology challenges, as deduced from the issues that SMEs trading in specific sectors are faced with and the mitigation support they required. Table 3 indicates the number of SMEs from each of the top 10 business sectors and the type of technical issues they have been supported with. Notably, data management and analytics support appear as the single most accessed area of technical assistance and expert knowledge, across all business sectors. This observation corroborates the principal aim of this study, of encouraging the use of data for SME growth. It can be argued that Table 3 also highlights training and skills development as a popular type of support, but it is worth noting that about one-fifth of the skills enhancement assistance itself was data-related.

---
[6] JavaScript Object Notation

## 4. DISCUSSION AND SOME LEARNED LESSONS

The adoption of data science techniques by SMEs requires some investment and also some time to take effect before they can expect to gain revenue from them. However, our study noted that most small businesses are reluctant to invest in methods and technologies if they cannot foresee quick profits. Statistically speaking and without excluding other potential applications, we have empirically identified four main current areas in which SMEs effectively use data analytics to grow their businesses.

Firstly, a common interest is the desire to improve digital marketing using data. In particular, SMEs derive immense benefits from the analysis of their data, for example identifying customer purchasing behaviour and patterns.

Secondly, SMEs use data to identify new markets through analysis of potential customer characteristics – e.g. using governmental open data, such as census data, demographics, detailed mapping available in the UK. Such analyses enable SMEs to locate markets and geographies where demand for their products and services would already be high, to enable them to move into new markets with reduced risk.

Thirdly, many SMEs use social media platforms for their digital marketing, and such tools generate data as customers interact with the SME through these platforms. Analysing this data proved useful to businesses in comparing their digital marketing performances across the different platforms (e.g., Facebook, Twitter, LinkedIn, and Instagram). The resulting insights are then used to focus on the best performing platform and allow the SME to make informed decisions about unsubscribing from those with poor performance, to reduce costs. The same analytics can also be used to identify what marketing strategies worked well based on specific KPIs, e.g., "likes", views, reach and impressions.

Finally, SMEs can perform explanatory analysis and investigate statistical properties and correlations of their past data with the objective of creating predictive models. This enables them to take advantage of cutting edge machine learning and artificial intelligence (AI) methods to, for example forecast the future demand for their products and services, and to automatically estimate production parameters.

In connection with the above, other lessons were also learned from the research collaborations in partnership with SMEs. In particular, we highlight next a number of issues, which were found to be the most common obstacles for the majority of SMEs, preventing them from fully adopting emerging technologies such as *data analytics* and *big data*.

### A. Perceived ease of conventional business styles

SMEs trade in specific business sectors and most always rely on conventional business methods. As such, many would be predisposed to avoid adopting disruptive technologies such as data analytics, and would prefer to continue their traditional business styles. This means that they are not tempted to make use of business data other than for record-keeping. One interpretation of this reluctance to take advantage of recent data usage advances is not only the limited understanding of the benefits for their businesses but also due to the perceived short-term disruption making such changes may bring – including learning new tools, buying and implementing software packages or paying for cloud computing, and hiring employees with data analytics skills. The outcome is however an overlooked data-driven business opportunity, which could have provided the SMEs with long-term benefits. One way of mitigating this phenomenon could be encouraging a business shift towards accepting a data-driven element in the SME's decision-making process.

### B. Low awareness of open data benefits

Open data is an increasingly growing idea which attempts to make data produced (often) by government departments public and freely available, for reasons of political transparency, economic value, or service and product improvement. Despite this, we found relatively low awareness amongst SMEs of the existence of such datasets and the value of related open data for their businesses. About 20% of the SMEs we supported did benefit from incorporating open data in education, e-recruitment, and transport in their businesses, helping them create new services or identify new markets. Thus, it could be argued that the more SMEs recognize the relevance and potential positive impact of open data, the more they can create innovative data-driven products and services at lower cost.

### C. Inadequate knowledge of data value and its potential

A high proportion of the SMEs we worked with were aware of the fact that data can be used not only for record-keeping but also to gain some useful insights about their business. However, SMEs have data science skills deficiencies, for instance, many reported that they did not know how to make the right choice of the appropriate data to be collected from their business operations for analysis. Limited knowledge in data handling tools was also indicated as another hurdle by some SMEs.

### D. Limited financial resources

Our research work with businesses found that many SMEs understand the potential and value of data and its analytics for the growth of their businesses. We also noted their willingness to "trial" data analytics and invest in related technologies to derive meaning from their data. However, their aspirations for adopting data technology are limited as they neither have the necessary data analytics knowledge and expertise, nor the budget needed to hire specialists or outsource their analytic needs. This is compounded by SMEs having limited access to funds and loans in comparison with the access enjoyed by big companies. This justifies the significance of ERDF projects which are intended to serve the purpose of bridging this gap, and highlights the need for even more financial and technical support to uplift the level of data analytics adoption by micro and small businesses.

### E. Shortage of available domain-specific data analysts

Analysing business data effectively often requires a complex blend of data analytics skills, as well as business context and domain-specific knowledge [5]. With very limited analysts in the job market who meet such criteria, the charges incurred when SMEs outsource their data analytics needs to experts, are comparatively huge. For some domains, e.g. additive manufacturing, the adoption of data science and machine learning for process control and modelling is in its infancy with very limited numbers of experts available who also understand the business context. The use of data for analytics and decision-making is correspondingly less common among manufacturing SMEs [16], which puts off many SMEs from the effective utilization of their data to enhance their growth and productivity.

*F. Inadequate up-to-date technical knowledge*

Developing and upgrading employee skills is crucial for the growth of small businesses. However, some of the barriers previously discussed, including the limited availability of finance, restrict SMEs from keeping pace with the dynamic and ever-changing technological advances in terms of employee skills and competence. We found that this is particularly applicable to smaller companies, e.g., the manufacturing and technology firms, where they face challenges in designing, developing, and testing new products or upgrading their existing ones for new purposes.

*G. Poor knowledge of available funds and data tools*

In this study we also observed an inadequate understanding of available funds among SMEs, particularly those which can be accessed to realize important innovative digital and data-driven ideas. Based on our assistance and collaboration with SMEs, we found the Knowledge Transfer Partnership (KTP) as one of the most appealing funding schemes due to the minimal contribution it requires from the SME and the high level of business impact which it can achieve. In addition, very few micro-businesses were found to be aware of some of the widely used data management and analytic tools, and the advantages their use could have for their business. Supporting such SMEs to adopt these tools (including Dropbox, PowerBI, and RStudio), proved crucial to many.

## 5. SUMMARY AND CONCLUSION

The recognition of data value for businesses and its use for decision making is growing among UK SMEs, as more tend to be using data and analytics to understand their customer purchasing patterns, performance of their digital marketing, and to embed AI aspects in their operations. However, SMEs seem to be far from being in full swing in the digital revolution due to their limited resources including the lack of enough finance to invest in IT infrastructure and to hire the right skilled experts. This paper presented brief analysis on digital and data usage trends and some lessons learned from a case study of 53 UK SMEs. Observed challenges, limiting SMEs from uncovering the hidden values of big data, are also discussed. In addition, we highlighted the significance of micro and small businesses to the UK economy in terms of their contribution to employment and GDP. We showed that interest in business data analytics has grown in the last ten years and continues to grow. Our analysis has also suggested that most SMEs collect and store some sort of business data but require skills to analyse and produce useful insights for data-driven decision making. This could mean that if SMEs are empowered with the right skills and/or supported financially for this purpose, they can make full utilization of data to help them, and hence the entire economy, grow.